\documentclass[10pt,aps,prl,amsmath,twocolumn,amssymb,nofootinbib]{revtex4-2}

\usepackage{orcidlink}
\usepackage{tensor}

\hypersetup{colorlinks,citecolor=blue}
\hypersetup{colorlinks=true,linkcolor=magenta,filecolor=magenta,    urlcolor=blue}
\begin{document}

\title{ Black holes with regular scalar hair in Brans--Dicke gravity via the Herglotz variational principle
}

\author{Marek Wazny}
\email{marek.wazny@stud.ubbcluj.ro}
\affiliation{Department of Physics, Babeș-Bolyai University, 1 Kogălniceanu Street, Cluj-Napoca 400084, Romania}

\begin{abstract}
Brans--Dicke theory is reformulated within the Herglotz variational principle (HVP), and an exact black hole solution with scalar hair is obtained for $\omega_{0}=0$ and vanishing potential $V(\phi)=0$. The scalar profile is strictly positive and the resulting stealth Schwarzschild solution arises without fixing the otherwise arbitrary Herglotz function $\eta(r)$. Motivated by the weak-field limit, the explicit choice $\eta(r)=\eta_{0}(r-2M)^k/r^{k+2}$, with $\eta_0$ a constant of dimension length, produces a scalar field configuration remaining regular at the black hole horizon. Consequently, the HVP provides a new mechanism for evading standard no-hair theorems in scalar-tensor theories.

\end{abstract}

\maketitle

\paragraph{Introduction.} The structure of black holes in general relativity (GR) is uniquely characterized by mass $M$, charge $Q$, and angular momentum $J$, as established by the classical no-hair theorems \cite{Penrose_1965, Hawking_1070}. In scalar-tensor theories, however, black holes may in principle support scalar hair. The prototypical member of this class was introduced by Brans and Dicke in 1961 \cite{Brans_1961}. Shortly thereafter, it was shown that Brans--Dicke theory does not admit nontrivial scalar hair \cite{bek1, bek2, Hawking:1971vc}. This result has since been extended to generalized Brans--Dicke models, in which the parameter $\omega$ depends on the scalar field \cite{Sotiriou_2012}, as well as to certain subclasses of Horndeski gravity \cite{bekenstein_1995, Hui_2013} (see \cite{review} for a review). To date, the only known solution in Brans--Dicke theory that evades these no-hair theorems is the Bocharova-Bronnikov-Melnikov-Bekenstein (BBMB) black hole \cite{Bocharova1970, Bekenstein1974, MTZ}. However, in this case the scalar field diverges at the event horizon, casting doubt on the physical interpretation of the solution.

The inability of (generalized) Brans--Dicke theory to support hairy black holes has motivated the exploration of more general scalar-tensor theories, including (beyond) Horndeski and Einstein-scalar-Gauss-Bonnet (EsGB) gravity \cite{Kobayashi_2019}. A variety of hairy black hole solutions have been discovered in both (beyond) Horndeski \cite{Babichev2014, bakopoulos2025}, and EsGB \cite{ Silva_2018, Doneva_2018, Antoniou2018} theories. In the former, such solutions arise from the introduction of many additional degrees of freedom (DOF), encoded in arbitrary functions of the scalar field and its derivatives. While these extra DOF expand the space of solutions, they also reduce the predictive power of the theory. The latter only introduces a single additional DOF in the form of a coupling function $f(\phi)$, to the Gauss-Bonnet invariant. 

Motivated by finding novel black hole solutions with minimal additional DOF, the simplest nonminimally coupled scalar-tensor theory (Brans--Dicke) is formulated within the Herglotz variational principle (HVP). Recently, it has been argued that any consistent coupling between the classical gravitational field and quantum matter is fundamentally irreversible \cite{Galley2023anyconsistent}, which hints at its dissipative nature. Since the HVP was historically developed to describe classical dissipative systems \cite{Herg1}, it provides a natural framework for such couplings. It generalizes the standard action principle by redefining the action through a differential equation, whose source is the Lagrangian. In the special case where the Lagrangian does not depend on the action, Hamilton's classical principle is recovered. The HVP also lacks thorough investigation, only recently being applied to GR \cite{Lazo_2017,Paiva_2022,gaset} and modified gravity theories \cite{Wazny_2025, Wazny_2026}. However, to date, no black hole solutions have been constructed within Herglotz-type gravity theories. Addressing this gap constitutes the main focus of the present work.

This paper will formulate the Brans--Dicke theory within the HVP and present the resulting metric and scalar field equations, focusing on the simplest case with $\omega_0=0$. The main purpose will be to show, in the case of vanishing potential $V(\phi)=0$, there exists a stealth Schwarzschild solution with nontrivial scalar profile that is everywhere positive. Although the Herglotz vector is arbitrary (similar to the coupling function in EsGB), the metric solution is independent of its exact form. However, by choosing a specific form of the Herglotz vector, motivated by the weak-field limit, the scalar field can remain regular at the horizon.
\paragraph{Herglotz-type Brans--Dicke theory.}
For a spacetime defined on an $n$-dimensional smooth manifold $\mathcal{M}$, with subset $\mathcal{V}$, boundary $\Omega$, and metric $g_{\mu\nu}$, the Herglotz variational problem is formulated as \cite{Paiva_2022}
\begin{align}\label{GRsys}
    \begin{split}
        \nabla_\mu s^\mu &= \mathcal{L}\left(g_{\alpha\beta}, \partial_\mu g_{\alpha\beta}, \phi, \partial_\mu\phi, s^\mu, x^\mu\right) ,\\
        S(\Omega) &= \int_\Omega d^{n-1}x\sqrt{h} \,n_\mu s^\mu = \int_\mathcal{V} d^nx \sqrt{-g} ~\nabla_\mu s^\mu,
    \end{split}
\end{align}
where $s^\mu$ is the differentiable action-density vector field, $h=\det h_{\mu\nu} $ and $g=\det g_{\mu\nu}$ are the determinants of the metrics on $\mathcal{V}$ and $\Omega$ respectively, $n_\mu$ is the unit vector normal to $\Omega$, and $\phi(x^\mu)$ is the Brans--Dicke scalar field.

Now define an arbitrary closed one form $\lambda = \lambda_\mu dx^\mu$, namely the Herglotz vector, which will couple to the action density $s^\mu$. Thus, the Lagrangian density (with $G=c=1)$ is given by\footnote{The focus of this work will be confined to the vacuum, hence the matter Lagrangian density is ignored for clarity.}
\begin{equation}\label{BDHLag}
    \mathcal{L} = \phi R -\frac{\omega_0}{\phi}(\nabla_\mu \phi \nabla^\mu \phi) -V(\phi) +\lambda_\mu s^\mu .
\end{equation}
where $\omega_0\in \mathbb{R}$ is the Brans--Dicke parameter, and $V(\phi)$ is the potential of the scalar field $\phi$. When taking the variation with respect to the metric in the HVP, only derivatives of the metric will pick up $\lambda_\mu$ contributions. The relevant variation is 
\begin{align}
&\int_\mathcal{V}d^n x\sqrt{-g} e^{-\psi} \phi\,g^{\mu\nu}\delta R_{\mu\nu} ,\quad \psi := \int \lambda_\mu dx^\mu .
\end{align}
 Computing this (for details see \cite{Paiva_2022}), in addition with the kinetic scalar term contribution, gives the field equations
 \begin{align}
     \phi G_{\mu\nu}+H_{\mu\nu } &= \frac{\omega_0}{\phi^2}\left(\nabla_\mu \phi\nabla_\nu \phi -\frac{1}{2}g_{\mu\nu} \nabla^\alpha\phi\nabla_\alpha \phi\right) \nonumber \\
     &\hspace{4.5mm}+ (\nabla_\mu \nabla_\nu -g_{\mu\nu}\Box) \phi -\frac{V}{2}g_{\mu\nu} ,
 \end{align}
 where $G_{\mu\nu}$ is the Einstein tensor and $H_{\mu\nu}$ is the Herglotz tensor defined by
\begin{align}
 H_{\mu\nu} &:= \phi K_{\mu \nu}+\lambda_\mu \partial_\nu \phi  +\lambda_\nu \partial_\mu \phi - 2g_{\mu\nu}\lambda^\rho \partial_\rho \phi , \\
 K_{\mu\nu} &:= \frac{1}{2}\left( \nabla_\mu\lambda_\nu + \nabla_\nu \lambda_\mu\right) -\lambda_\mu \lambda_\nu - g_{\mu\nu}\left(\nabla_\rho \lambda^\rho-\lambda_\rho \lambda^\rho\right) .
\end{align}
For the scalar field one finds the equation of motion 
\begin{equation}
    \frac{2\omega_0}{\phi}(\nabla_\mu -\lambda_\mu)\nabla^\mu\phi-\frac{\omega_0}{\phi^2}\nabla_\mu\phi \nabla^\mu \phi +R- V'=0 ,
\end{equation}
where $V' := dV/d\phi$.

The aim of this work is to demonstrate that the simplest nonminimally coupled scalar-tensor theory formulated within the HVP is sufficient to yield new and interesting results. As such, $\omega_0=0$ will be adopted henceforth. This leads to the metric field equations
\begin{align}\label{Fieldeq}
    \mathcal{E}_{\mu\nu}:= \phi G_{\mu \nu} +H_{\mu \nu} + \left(  g_{\mu \nu} \Box - \nabla_{\mu} \nabla_{\nu} \right)\phi+ \frac{V}{2}g_{\mu\nu} = 0    ,
\end{align}
and the scalar field equation
\begin{align}
    \mathcal{S} := R - V' =0 .\label{onshell} 
\end{align}
The trace of Eq.~(\ref{Fieldeq}) will also prove useful yielding
\begin{equation}\label{trace}
    \phi R - 3\Box\phi-H -2V =0 ,
\end{equation}
where $H := \tensor{H}{^\mu _\mu}$. This allows the scalar field equation to take the more canonical form
\begin{equation}\label{scalareom}
    3\Box\phi +H= \phi V'-2V .
\end{equation}
It also allows Eq.~(\ref{Fieldeq}) to be written in the traceless form
\begin{align}\label{tracelessfield}
    &\phi\left(R_{\mu\nu} -\frac{1}{4}Rg_{\mu\nu}\right) + \left(\frac{1}{4}g_{\mu\nu}\Box-\nabla_\mu\nabla_\nu\right)\phi \nonumber\\
    &=\frac{1}{4}Hg_{\mu\nu}-H_{\mu\nu}.
\end{align}

Notice that taking the divergence of Eq.~(\ref{Fieldeq}) leads to
\begin{equation}
    \nabla^\mu \mathcal{E}_{\mu\nu} = \frac{\mathcal{S}}{2}\nabla_\nu \phi -\nabla^\mu H_{\mu\nu} .
\end{equation}
This stands in direct contrast with general scalar-tensor theories, in which any solution of the metric field equations automatically satisfies the scalar field equation, that is, $\mathcal{E}_{\mu\nu}=0  \Rightarrow \mathcal{S}=0$. In the HVP, however, this implication does not hold. This is a direct consequence of the dissipative nature associated with the HVP. One must therefore verify explicitly that either $\nabla_\mu H_{\mu\nu}=0$, Eq.~(\ref{onshell}), or Eq.~(\ref{scalareom}) is satisfied in order to confirm the consistency of any solution obtained from the field equations.

\paragraph{Weak-field limit.}
To obtain the weak-field limit one takes the usual ansatz
\begin{equation}
    ds^2 = -(1+2\Phi)dt^2 + (1-2\Phi)(dr^2+r^2d\Omega^2) ,
\end{equation}
where $\Phi$ is now the modified Newtonian potential and $d\Omega^2_2$ is the metric on the unit 2-sphere, $S^2$. With the addition of the Brans--Dicke and Herglotz fields the usual Laplace equation for the Newtonian potential now gives a Poisson equation for the modified potential $\Phi$. This is found in the typical manner by the $(t,t)$ component of the field equations
\begin{equation}\label{35}
   G_{tt} = -\frac{1}{\phi}\left[g_{tt}\left(\Box\phi +\frac{V}{2}\right) +\tensor{\Gamma}{^r _{tt}}\partial_r\phi +H_{tt}\right].
\end{equation}
To proceed further it can be noted that the maximally symmetric nature of the spacetime constrains the Herglotz field to only have a radial component with radial dependence, that is $\lambda^\mu = (0,\eta(r),0,0)$, for some scalar function $\eta(r)$. Similarly, $\phi(r)$ can only have radial dependence. Then, working to first order in each field, Eq.~(\ref{35}) reads
\begin{align}\label{Phigen}
    \frac{1}{r}\partial_r^2(r\Phi) &= -\frac{1}{2\phi}\Bigg[(1+2\Phi)\left(\frac{\phi}{r^2}\partial_r(r^2\eta) +2\eta\partial_r\phi-\frac{V}{2}\right) \nonumber\\
    &\hspace{4.5mm}-3\phi\eta\partial_r\Phi -\frac{(1+4\Phi)}{r}\partial_r^2(r\phi) +\partial_r\Phi \partial_r\phi \Bigg]  .
\end{align}

To solve this equation some assumptions must be made. In the weak-field limit $\phi$ admits a Taylor expansion which at first order gives $\phi \approx$ constant. Then, if the modified potential is restricted such that it reproduces the Newtonian potential $\Phi=\Phi_N:= -M/r$, where $M$ is the mass of the system, Eq.~(\ref{Phigen}) reduces to
\begin{equation}\label{etasol1}
    \frac{1}{\eta} \partial_r\eta = \frac{7M-2r}{r(r-2M)},
\end{equation}
for a trivial potential $V=0$. Similarly, if one takes the $(t,t)$ component of the traceless equation, Eq.~(\ref{tracelessfield}), the above computations yield
\begin{equation}\label{etasol2}
    \frac{1}{\eta}\partial_r\eta = \frac{10M-2r}{r(r-2M)},
\end{equation}
which has the advantage of being independent of the potential $V$. The solution is then given by 
\begin{equation}\label{etasol}
    \eta(r) = \eta_0\frac{(r-2M)^k}{r^{k+2}} ,
\end{equation}
for integration constant $\eta_0$, where $k=3/2$ for Eq.~(\ref{etasol1}) and $k=3$ for Eq.~(\ref{etasol2}).

\paragraph{Stealth Schwarzschild solution.}
For the sake of simplicity, a spherically symmetric and static metric is imposed, leading to the ansatz
\begin{equation}\label{staticds}
        ds^2 = -f(r)dt^2 +\frac{1}{h(r)}dr^2 +r^2 d\Omega_2^2 ,
\end{equation}
for scalar functions $f(r)$ and $h(r)$. In order to obey the symmetries of the metric, the Herglotz vector reduces to 
\begin{equation}\label{eta}
    \lambda^\mu = (0, \eta(r), 0, 0),\quad \lambda_\mu = \left(0,\frac{\eta(r)}{h(r)},0,0\right) ,
\end{equation}
for some scalar function $\eta(r)$. The Brans--Dicke scalar also becomes a function of radial coordinate $\phi(r)$.
To solve the field equations, substitute Eqs.~(\ref{staticds})\,-\,(\ref{eta}) into Eq.~(\ref{Fieldeq}) along with the assumptions outlined above, which yields
\begin{align}
    \mathcal{E}_{tt} &=  \phi G_{tt}-f\left(\Box\phi +\frac{V}{2}\right) +\frac{1}{2}h\partial_rf\partial_r\phi\nonumber\\
    &\hspace{4.5mm}+f\phi\left[\partial_r\eta+\eta\left(-\frac{1}{2}\frac{\partial_rh}{h}+2 \frac{\partial_r\phi}{\phi}-\frac{\eta}{h}+\frac{2}{r} \right) \right]   \label{tt} ,\\
    \mathcal{E}_{rr} &= \phi G_{rr}+\frac{1}{h}\left(\Box\phi+\frac{V}{2}\right) -\partial_r^2\phi -\frac{1}{2}\frac{\partial_rh}{h}\partial_r\phi \nonumber\\
    &\hspace{4.5mm}-\frac{\eta\phi}{h}\left(\frac{1}{2}\frac{\partial_r f}{f}+\frac{2}{r}\right)  \label{rr} ,\\
    \mathcal{E}_{\theta\theta} &= \phi G_{\theta\theta}+ r^2\left(\Box\phi +\frac{V}{2}\right)-rh\partial_r\phi \nonumber\\
    &\hspace{2.5mm}-r^2\phi\left[\partial_r\eta+\eta\left(\frac{1}{2}\frac{\partial_rf}{f}-\frac{1}{2}\frac{\partial_rh}{h}+2 \frac{\partial_r\phi}{\phi}-\frac{\eta}{h}+\frac{1}{r} \right) \right]  \label{hh},
    \end{align}
where 
\begin{equation}
\Box\phi = h\left[\partial_r^2\phi+ \left(\frac{1}{2}\frac{\partial_rh}{h}+\frac{1}{2}\frac{\partial_rf}{f}+\frac{2}{r}\right)\partial_r\phi\right] ,
\end{equation}
and noting $\mathcal{E}_{\theta\theta} = \mathcal{E}_{\varphi\varphi}r^2\sin(\theta)$. The scalar field equation takes the form Eq.~(\ref{onshell}) or Eq.~(\ref{scalareom}) with the latter depending on the trace of the Herglotz tensor 
\begin{equation}
     H = -3\phi \left[\partial_r\eta +\eta\left(\frac{1}{2}\frac{\partial_r f}{f}-\frac{1}{2}\frac{\partial_r h}{h} + 2\frac{\partial_r\phi}{\phi}-\frac{\eta}{h} +\frac{2}{r}\right) \right] .
\end{equation}

In the case $V=0$, the scalar equation takes the simple form $R=0$, reducing the Einstein tensor to the Ricci tensor. For homogeneous solutions $f=h$, this leads to the solution
\begin{equation}\label{fgen}
    f(r) = 1+\frac{C_1}{r}+\frac{C_2}{r^2} ,
\end{equation}
for integration constants $C_1$ and $C_2$. Then, note that each of Eqs.~(\ref{tt})\,-\,(\ref{hh}) vanish independently so the linear combination $\frac{1}{f^2}\mathcal{E}_{tt}+ \mathcal{E}_{rr}+\frac{1}{r^2f}\mathcal{E}_{\theta\theta}=0$ may be taken to yield the constraint
\begin{align}\label{etav0}
    \eta(r) &= \frac{r^3(C_1 +r) -C_2(C_2  +C_1r)}{r^2 (r^2-C_2)} \frac{\partial_r \phi}{\phi} \nonumber \\
    &\hspace{4.5mm}+\frac{C_2(C_2 + C_1r + r^2)}{r^3 (r^2-C_2)} .
\end{align}
In order for each of Eqs.~(\ref{tt})\,-\,(\ref{hh}) to be satisfied after substituting Eq.~(\ref{fgen}) and Eq.~(\ref{etav0}), one obtains the condition $\phi=0$ or $C_2=0$. To have a nontrivial scalar the latter must be chosen. Furthermore, the case $\phi=0$ is fundamentally distinct from $\phi=$ constant and is not only mathematically degenerate but also physically inadmissible since the usual interpretation of an effective gravitational coupling strength $G_{\text{eff}}\sim \phi^{-1}$ becomes singular.

Identifying $C_1=-2M$ gives the metric solution
\begin{equation}\label{SS}
    ds^2 = -\left(1-\frac{2M}{r}\right)dt^2 + \left(1-\frac{2M}{r}\right)^{-1}dr^2 +r^2d\Omega_2^2 ,
\end{equation}
along with the Brans--Dicke scalar
\begin{equation}\label{phisol}
    \phi(r) =  \phi_0\exp\left(\int dr\,\frac{r}{r-2M}\eta(r)\right) ,
\end{equation}
where $M$ is the ADM mass, and $\phi_0$ is an arbitrary integration constant.

This corresponds to a stealth Ricci-flat solution with a class of non-trivial scalar profiles depending on $\eta(r)$, similar to what appears in the Bocharova--Bronnikov--Melnikov--Bekenstein (BBMB) black hole. The key difference in this case being the form of the scalar field, and the metric being exactly Schwarzschild. Consequently, the no hair theorems of \cite{Hawking:1971vc} and \cite{Sotiriou_2012} would normally apply in the traditional Lagrangian variational principle. Indeed, in the case $\eta=0$, the scalar field $\phi$ reduces to a constant and the usual case with trivial scalar hair is recovered. However, within the HVP the scalar field $\phi$ is strictly greater than zero but is only unique up to the arbitrary function $\eta(r)$. This is expected since the HVP introduces an additional, non-dynamical degree of freedom which naturally supports the scalar configuration. To reduce this arbitrariness one can impose physical, boundary, and regularity conditions on $\eta(r)$. 

To show this explicitly, a physically motivated choice is Eq.~(\ref{etasol}), since it was shown to be compatible with the weak-field limit. Choosing this form also provides the desired boundary condition immediately, in particular, it decays to zero at spatial infinity. The regularity condition explicitly effects the regularity of $\phi(r)$ and depends on the constant $k$. Previously, the value $k$ in this solution took on discrete values, however in the following the analysis will be generalized to any positive value. For a horizon at $r=2M$ the solution for $\eta$ leads to the scalar profile
 \begin{equation}\label{phisol2}
     \phi(r) = \phi_0\exp\bigg\{\frac{\eta_0}{2Mk}\left(1-\frac{2M}{r}\right)^k\bigg\}.
 \end{equation}
This solution tends to a constant at infinity and behaves near the horizon as
\begin{equation}
    \phi(r) \sim (r-2M)^k ,
\end{equation}
hence, the regularity dependence on $k$ becomes apparent at the horizon. Discussing cases on general $k>1$, one finds for integer $k$, the solution for $\phi$ is regular and $C^{\infty}$ at the horizon. For non-integer $k$ where $k<p\in \mathbb{N}^{>1}$, $\phi$ remains regular at the horizon but is at most $ C^{p-1}$, with its $p$th derivative diverging at the horizon.

One may also consider the stability of the solution Eq.~(\ref{phisol}) with a fixed Herglotz vector and background metric. Focusing on Eq.~(\ref{scalareom}) and perturbing the solution $\phi=\bar\phi+\delta\phi$, one finds 
\begin{align}\label{pert1}
    0&=3\bar\Box(\bar\phi+\delta\phi)+\delta H \nonumber\\
    &= \bar\Box\delta\phi  -\left(\partial_r\bar\eta  -\frac{\bar\eta^2}{\bar f} +\frac{2\bar\eta}{r} \right)\delta\phi -2\bar\eta \partial_r \delta\phi ,
\end{align}
where barred quantities are the solutions leading to Eq.~(\ref{SS}). Decomposing $\delta\phi$ as $\delta\phi = e^{-i\omega t} \psi(r) Y_{lm}(\theta,\varphi)/r $, where $Y_{lm}$ are the standard spherical harmonics, Eq.~(\ref{pert1}) becomes
\begin{align}
      \frac{d^2\psi}{dr_*^2} -2\bar \eta\frac{d\psi}{dr_*} +\left[\omega^2-V_{\text{eff}}\right]\psi=0 ,
\end{align}
for the radial coordinate $ dr = \bar f dr_* $ and effective potential
\begin{equation}
    V_{\text{eff}}(r) := \bar f\left(\frac{l(l+1)}{r^2}+\frac{\partial_r \bar f}{r} +\partial_r \bar\eta -\frac{\bar\eta^2}{\bar f}\right).
\end{equation}
Even after the standard radial coordinate redefinition, a first order derivative still appears, which is not surprising given the dissipative nature of the HVP. If this term were positive it would help stabilize the perturbations, but if $\bar \eta$ decays fast enough, such as $\bar\eta\sim 1/r^2$, then this term becomes irrelevant away from the horizon. In this case, the effective mass, $\mu_{\text{eff}} := \partial_r\bar\eta -\bar\eta^2/\bar f$, controls the stability. Thus, this framework also accommodates the possibility of tachyonic instabilities by satisfying the sufficient condition \cite{Buell}

\begin{equation}
    \int_{2M}^\infty dr \frac{V_{\text{eff}}(r)}{1-\frac{2M}{r}} <0 .
\end{equation}
For the specific form Eq. (\ref{etasol}), this leads to the condition
\begin{equation}
    M^2 < \frac{\eta_0}{2(k+1)(2k+1)(2k+3)} ,
\end{equation}
for $l=0$ modes.

It is beyond the scope of this paper, but a complete characterization of stability would require perturbing the metric as well. One may anticipate that the same dissipative term would appear in the full Regge–Wheeler–Zerilli equations, however it is not clear if they would decouple as in the standard GR case. Such an analysis would also clarify whether distinct choices of the Herglotz function $\eta(r)$ give rise to genuinely different scalar hairs, or simply create a gauge equivalent class as suggested by the fact that the stealth metric remains unaffected by $\eta(r)$. 

 \paragraph{Conclusions.}
In the present work, a stealth Schwarzschild solution with nontrivial scalar hair has been obtained in a Herglotz-type Brans--Dicke theory with $\omega_0=0$ and vanishing potential $V(\phi)=0$. Although the Herglotz vector is arbitrary, only the regularity of the scalar field solution depends on its explicit form. The choice $\eta(r)=\eta_{0}(r-2M)^k/r^{k+2}$ is physically motivated by the weak-field limit and yields a scalar configuration that remains regular at the black hole horizon for certain choices of $k$. This presents the HVP as a method for evading scalar no-hair theorems in Brans--Dicke and related scalar-tensor theories. It also provides a natural dissipative setting for the irreversibility between quantum matter and the classical gravitational field.

Another established method for circumventing no-hair theorems appears in the scalarization of EsGB theory. However, within the HVP, the Gauss-Bonnet invariant contributes to the field equations without requiring an arbitrary coupling function, suggesting that a Herglotz-Einstein-Gauss-Bonnet theory may offer a promising direction for further investigation.

It would also be interesting to explore non-vanishing scalar potentials, such as $V \sim m^{2}\phi^{2}$, which may yield solutions carrying primary scalar via a scalar mass $m$. Allowing $\omega_{0}\neq 0$ may similarly broaden the solution space and introduce extra dynamics. For example, by adding a Maxwell contribution to the current Lagrangian, the Herglotz formalism alters its equation of motion, leading to the non-zero components $F_{tr} =-F_{rt}= Q_{\text{eff}}(r)/r^2$ for a field with electric charge $Q$, where 
\begin{equation}
    Q_{\text{eff}}(r) =Q\sqrt{\frac{h}{f}}\exp\left(\int dr \frac{\eta(r)}{h(r)}\right) ,
\end{equation} is a radially dependent effective electric charge. A surface level analysis suggests even for homogeneous solutions $f=h$, a Reissner--Nordström (RN) solution would not be constructable since the most general metric form given by the scalar equation $R=0$ can only support constant electric charge. Therefore, additional dynamics via a non-zero potential or coupling parameter would seemingly need to appear for RN geometries. However, a more detailed analysis could suggest any Herglotz-type Brans--Dicke black hole is unable to support electric charge, which would align with the fact that realistic black holes would likely be electrically neutral. 

Furthermore, astrophysically relevant rotating spacetimes remain unexplored; extending the current framework to Kerr geometries represents another important step toward assessing the physical relevance of Herglotz-type scalar-tensor theories. Given the dissipative nature of the HVP, one could explore whether this framework provides a mechanism to limit the angular momentum of rotating black holes. Such an effect would lead to observable signatures, particularly when combined with a full stability analysis that may yield quasinormal modes (QNMs) differing from those predicted by GR.

\paragraph{Acknowledgments.}
The author would like to thank both Lehel Csillag and Daniela Doneva for the valuable discussions.

\bibliography{refs}

\end{document}